\documentclass[a4paper,10pt,reprint,aps,pre,superscriptaddress]{revtex4-1}
\usepackage[utf8]{inputenc}
\usepackage{amsmath,amsfonts,amssymb}
\usepackage[utf8]{inputenc}
\usepackage{graphicx}
\usepackage{color}
\usepackage{xcolor}
\definecolor{redcolor}{rgb}{.0,0.,0.}

\begin{document}

\title{Stochastic dynamics and the predictability of big hits in online videos} 

\author{Jos\'e M. Miotto}
\affiliation{
Max Planck Institute for the Physics of Complex Systems, 01187 Dresden, Germany
}

\author{Holger Kantz}
\affiliation{
Max Planck Institute for the Physics of Complex Systems, 01187 Dresden, Germany
}

\author{Eduardo G. Altmann}
\affiliation{
Max Planck Institute for the Physics of Complex Systems, 01187 Dresden, Germany
}
\affiliation{
School of Mathematics and Statistics, University of Sydney, NSW 2006, Australia
}

\date{\today}

\begin{abstract}
The competition for the attention of users is a central element of the 
Internet. 
Crucial issues are the origin and predictability of big hits, the few items 
that capture a big portion of the total attention. 
We address these issues analyzing 10 million time series of videos' views 
from YouTube.
We find that the average gain of views is linearly proportional to the number 
of views a video already has, in agreement with usual rich-get-richer 
mechanisms and Gibrat's law, but this fails to explain the prevalence of big 
hits. 
The reason is that the fluctuations around the average views are themselves 
heavy tailed.
Based on these empirical observations, we propose a stochastic differential 
equation with L\'evy noise as a model of the dynamics of videos. 
We show how this model is substantially better in estimating the 
probability of an ordinary item becoming a big hit, which is considerably 
underestimated in the traditional proportional-growth models.
\end{abstract}

\maketitle

\section{Introduction.}
YouTube is a representative example of online platforms in which 
items (videos in this case) compete for the attention of 
users~\cite{simon1971designing,wu2007novelty,salganik2006experimental}.
The popularity of videos vary by orders of magnitude, resembling the fat-tailed 
distributions that have been reported in other online 
systems~\cite{huberman2009crowdsourcing,miotto2014predictability,
weng2012competition}, in income and wealth~\cite{pareto1964cours}, in
finance~\cite{mandelbrot1963prices}, and in 
disciplines such as ecology, earth science, and physics~\cite{Newman2009}.
The origin of such fat-tailed distributions is a century-old problem that lies 
at the heart of complex-systems 
science~\cite{yule1925mathematical,gibrat1931inegalites,mandelbrot1960pareto,
simon1958size,Simkin2011,perc2014matthew}.
At the core of the different proposed models lies the idea that the current 
popularity (wealth) determines the future popularity gain (income) and enhances 
the inequality (rich-get-richer).
Indeed, such (linear) {\it proportional growth} is the essential ingredient 
of Gibrat's law (used to describe the growth of 
firms~\cite{gibrat1931inegalites,simon1958size} and 
cities~\cite{gabaix1999zipf}), the Yule-Simon model (to model 
species genera~\cite{yule1925mathematical} and 
language~\cite{simon1955class,gerlach2013stochastic}), scientific 
memes~\cite{kuhn2014inheritance} and the preferential attachment model of 
network growth~\cite{barabasi1999emergence}. 
Proportional growth suggests that the big hits are very predictable because 
they originate from early advantages that are amplified over time.

\begin{figure*}[htbp]
\begin{center}
\includegraphics[width=8.6cm]{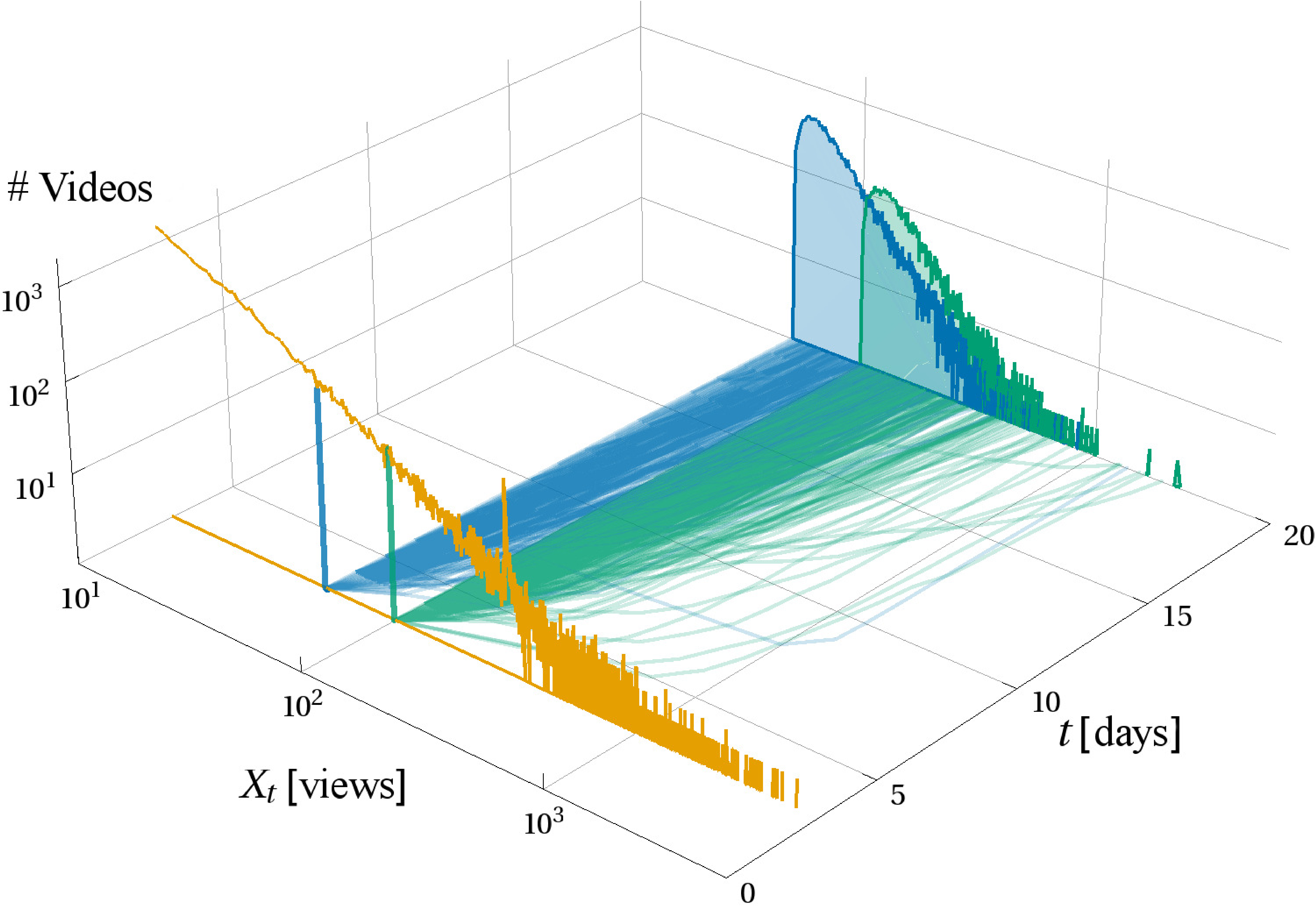}
\includegraphics[width=8.8cm]{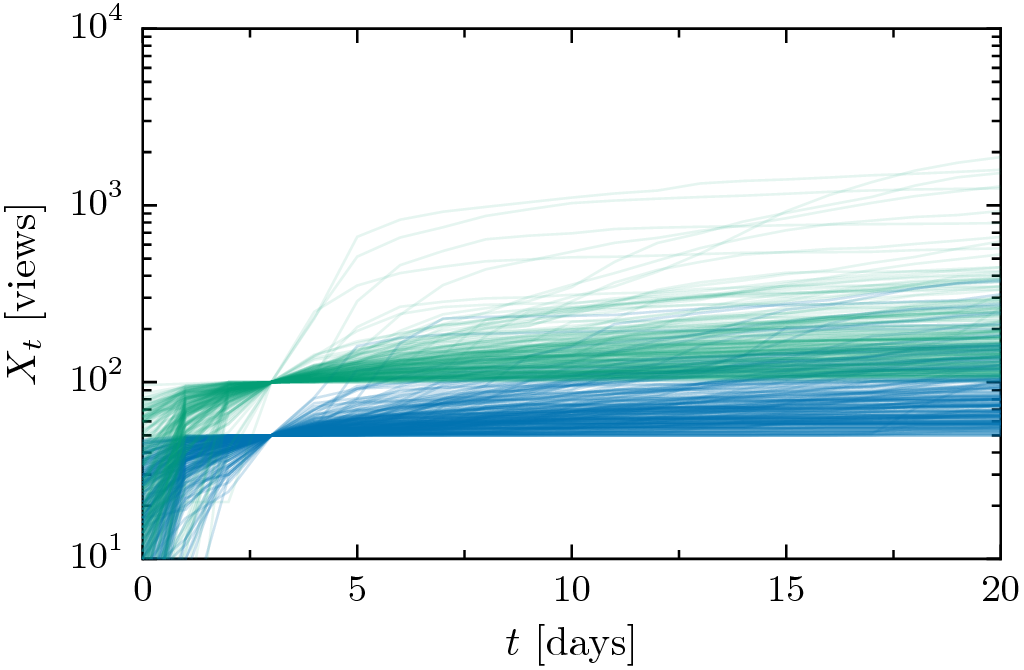}
\caption{ Evolution of videos' views $X_t$ as a function of the time $t$ after 
  publication. After $t=3$ days the distribution of views is already heavy-tailed (orange histogram).
Videos having initially the same amount of  views show very distinct evolutions. This is illustrated here by highlighting two group of videos with the same number of views at $t=3$ (blue $X_{t=3}=50$ 
  and green $X_{t=3}=100$). Each line (at the bottom and in the right plot) corresponds to the trajectory of one video. On the back, the histograms of the two groups of videos at $t=20$ days are shown.
  }
\label{fig.1}
\end{center}
\end{figure*}

The application of growth models to describe the popularity of online items 
bring new opportunities and challenges. 
On the one hand, due to the increasing availability of datasets, it becomes 
possible to compare models with an unprecedent accuracy.
On the other hand, the expectations we have of the models are higher. 
For instance, a central question is to forecast and identify the origins of the 
big hits~\cite{wang2013quantifying,WattsBook}, the most successful videos which 
capture most of the attention and produce most of the revenue through 
advertisement.
To address this and other questions, the characterization of the heavy-tailed 
distribution of aggregated activity is not enough.
One has to: (i) improve the description of the dynamics of individual items; 
and 
(ii) go beyond the average growth and analyze the stochastic 
fluctuations~\cite{maillart2008empirical,ratkiewicz2010characterizing}.
The importance of these factors is illustrated in Fig.~\ref{fig.1}, where we 
show trajectories (views vs. time) of videos with the same early success (the 
same number of views, 3 days after publication).
We see that trajectories quickly spread and that many trajectories with a weak 
start become popular in time. This suggests that big hits have a low 
predictability (i.e., they are hard to anticipate).

In this article, we investigate the predictability of big hits using stochastic 
models of individual items. Predictability is the possibility of 
anticipating the future based on present information and we confront the
predictability expected from models to observations in the data.
We compare traditional growth models to data ($X_t$, views over time) of 
more than 10 million YouTube videos.
We find that previously proposed models are unable to correctly account for the 
(random) fluctuations observed in the data, which we find to be described 
by a L\'evy-stable distribution. 
We propose and validate a stochastic model that explains such reduced 
predictability by incorporating both proportional growth and L\'evy noise. 
This shows that, even if present, proportional growth is not the only 
responsible for the origin of fat-tailed distributions.
Finally we show that our model substantially improves the prediction of the 
probability of big hits, but that unexpected big hits have an even higher 
probability in real data due to temporal correlations not accounted by this 
class of models.

\section{Theoretical Framework}
YouTube is a website where videos generated by third parties are shared. 
It is the third most visited website of all Internet.
We collected more than 10 million time series of the daily number of views of 
videos published between Dec 2011 and Mar 2013~\cite{miotto2014timeseries}.
The number of views a video receives depends on the interplay between its 
content and various factors.
Videos related to ongoing events are strongly influenced by their development, 
its media coverage, and other factors exogenous to the online activity of users.
Videos are also influenced by endogenous factors, such as the sharing and 
recommendation in online media, generating cascades of activity in the 
social 
network~\cite{sornette2004endogenous,watts2002simple}.
Additionally, a video can be viewed by following a link from a related video, 
i.e. hopping through the videos' network which changes continuously according 
to YouTube's recommendation and promotion algorithms.
The interplay and feedback between these and other factors lead to the complex 
dynamics we observe in the time series.
Modeling specific factors~\cite{chatzopoulou2010first} and differentiating 
between them (e.g., between exogenous and endogenous
factors~\cite{crane2008robust,
sornette2004endogenous,ghanbarnejad2014extracting}) are topics of recent 
research.
This approach is difficult to be pursued because it requires detailed 
information of user activities and the possibility of isolating the factors.
Instead, here we aim at a coarse-grained description of the dynamics of 
attention in which the combination of the different factors described above are 
{\it effectively} accounted by deterministic and stochastic terms.

Let $X_t$ be the cumulative number of views that a video received in the first 
$t$ days after its release.
A very general stochastic model for the growth of $X_t$ in $t$ is the diffusion 
process~\cite{risken1984fokker,maillart2008empirical,mollgard2015emergent}
\begin{equation}\label{eq.sde1}
  dX_t = \mu(t,X_t)dt + \sigma(t,X_t) dW_t,
\end{equation}
where $W_t$ is a Wiener process ($\langle W_t\rangle=0$ and 
$\langle W_t^2\rangle=dt$), $\mu(t,X_t)$ is the average growth, and 
$\sigma(t,X_t)$ scales the fluctuations; an additional cutoff in $dW_t$ is 
added to ensure that $d X_t >0$.
We consider all videos to be indistinguishable so that variations in the 
behavior of videos with the same $X_t$ should be accounted by the 
stochastic term $\sigma(t,X_t) dW_t$.
Extensions of our model could consider $\mu(t,X_t)$ and $\sigma(t,X_t)$ to 
depend on properties of the video and on $X_{t'}$ for $t'<t$.

\section{Data analysis}
We now analyze the data in order to identify the functions $\mu(t,X_t)$ 
and $\sigma(t,X_t)$.
Since the minimum resolution of our data is $\Delta t=1$ day, the 
models we propose aim to fit the quantity $\Delta X_t=X_{t+1}-X_t$, 
the number of views obtained exclusively in the day $t+1$.
We first focus on the deterministic term of Eq.~(\ref{eq.sde1}), $\mu(t,X_t)$. 
In linear proportional growth models the average growth is proportional to the 
views, $\mu(t,X_t) = \mu_t X_t$, where the temporal dependence on $\mu_t$ 
accounts for the decay in the attention gain~\cite{wu2007novelty};
this decay is very strong in the first weeks, so we will focus on the days 
up to $t=30$.

\begin{figure*}[!ht]
\begin{center}
\includegraphics[width=17.4cm]{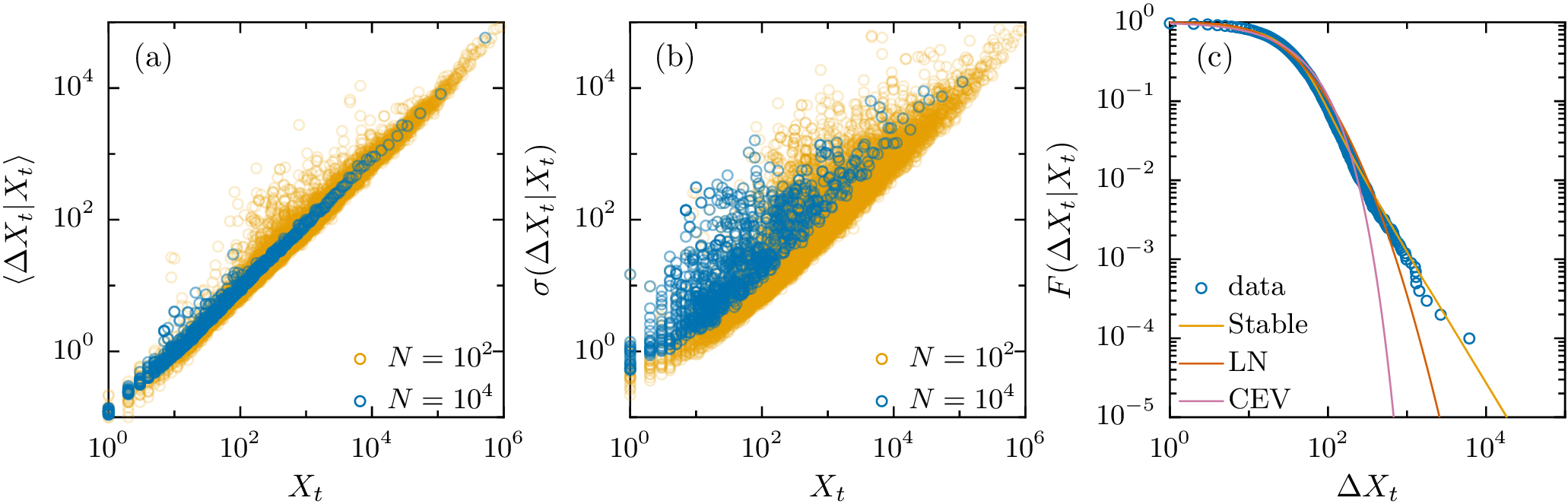}
\caption{
  Average and fluctuations in the growth of YouTube Videos.
  \textbf{(a)} Mean $\langle \Delta X_t \rangle$ and 
  \textbf{(b)} standard deviation $\sigma(\Delta X_t$) for videos with 
$X_t$ views $t=3$ days after publication.
  Both $\langle \ldots \rangle$ and $\sigma$ are computed  in windows centered 
  at $X_t$ and containing $N$ items (see legend).
  \textbf{(c)} Complementary cumulative density function 
  $F(\Delta X_t | X_t)$ for $X_t\in[499,513]$ and $t=3$.
  Data (blue circles) is compared with fits of three distributions 
  (S: L\'evy-stable, LN: Lognormal, CEV: Constant Elasticity of 
  Variance) and confirms the existence of heavy tails which are best 
  described by the L\'evy distribution.
  }
\label{fig.2}
\end{center}
\end{figure*}

This condition is consistent with our data: in Fig.~\ref{fig.2}(a) we see for a 
fixed $t$, that the dependence of the conditional average $\langle \Delta 
X_t| X_t\rangle$ (computed in windows of 
$N$ videos) with $X_t$ is roughly a line with slope $1$, a standard method to 
check for proportional growth \cite{newman2001clustering,perc2014matthew}.
We now repeat the analysis for the stochastic term $\sigma(X_t,t)$ of 
Eq.~(\ref{eq.sde1}).
A natural proposal for $\sigma$ is $\sigma(X_t,t) = \sigma_t X_t^\beta$~\cite{mollgard2015emergent}, where the $\beta$ parameter allows us to 
model a possible fluctuations' scaling, in the form of the Taylor's 
Law~\cite{eisler2008fluctuation}.
In particular, the $\beta=1$ case used in Ref.~\cite{maillart2008empirical}, is 
equivalent to $Y_t=\ln X_t$ exhibiting constant fluctuations, and corresponds 
to a Geometric Brownian Motion.
The simplest way to evaluate the stochastic term in this context is to repeat 
what was done for the mean and measure the standard deviation $\sigma$ in 
a window of $N$ items centered around $X_t$~\cite{maillart2008empirical}.
This is equivalent to the standard estimation of the drift and diffusion 
coefficients in a Fokker-Planck Equation~\cite{friedrich1997description}.
Results in Fig.~\ref{fig.2}(b) confirm the roughly linear scaling in the double 
logarithmic scale, in agreement with $\sigma(X_t,t) \propto X_t^\beta$ with 
$\beta \approx 1$.
However, in opposite to the case of $\mu(X_t,t)$ shown in panel (a), the data 
show strong fluctuations across $X_t$ and depend on the sample size $N$ (the 
larger the $N$ the larger the measured $\sigma$).

The observations above motivate us to look at the full probability 
distribution $\mathbb{P}(\Delta X_t|X_t)$~\cite{amaral1998power}.
In Fig.~\ref{fig.2}(c) we see in the particular histogram 
$\mathbb{P}(\Delta X_3|X_3 
\approx 500)$, that the distribution has a heavy tail; this explains the 
observation that $\sigma$ grows with $N$, i.e. 
$\mathbb{P}(\Delta X_t|X_t)$ has a 
diverging second moment~\cite{bouchaud1990anomalous}.
Heavy-tailed fluctuations of $\Delta X_t$ may still be compatible with 
Eq.~(\ref{eq.sde1}) if one considers that the temporal interval used in our 
analysis is not infinitesimal $\Delta t=1$ day $\gg dt$; in deed, Gaussian 
fluctuations are expected only when $\Delta t \rightarrow 0$.
In this case, the stochastic differential equation has to be integrated 
up to $\Delta t$, so the 
fluctuations predicted from Eq.~(\ref{eq.sde1}) can be Lognormal (for 
$\beta=1$) or a distribution arising from the Constant Elasticity of Variance 
model (CEV, for $\beta\neq1$)~\cite{chen2010cev}, as shown in App.~\ref{app.1}.
Besides Eq.~(\ref{eq.sde1}), classical models associated with Gibrat's law 
(Champernowne-Gabaix or Yule-Simon) predict 
$\mathbb{P}(\Delta X_t|X_t)$ to have 
either short tails or Lognormal distributions (see App.~\ref{app.2}).
Beyond the Lognormal and CEV distributions, which follow from 
Eq.~(\ref{eq.sde1}), we consider also the L\'evy-stable 
distribution (S) 
because it originates from the generalized Central Limit Theorem for variables 
without finite variance~\cite{vm1986one}.
In Fig.~\ref{fig.2}(c) we show the fits of discretized versions of these 3 
distributions to the particular histogram discussed above.
The best fit is obtained by the (completely asymmetric) L\'evy-stable 
distribution (with a difference in the Bayes Information 
Criterion~\cite{schwarz1978estimating,noteBIC}, BIC, of 178 and 175, with respect to 
the Lognormal and CEV models). 
This result, which is confirmed below for different $X_t$ and $t$, indicates
that the fluctuations observed in the data are not compatible with the Wiener
process $W_t$ in Eq.~(\ref{eq.sde1}), and that the analysis of the mean 
and standard deviation done for Fig.~\ref{fig.2}(a) and (b) may be not enough 
to define the functions of Eq.~\ref{eq.sde1}.

\section{Alternative Model}
Motivated by the better fit of the L\'evy distribution and by the linear 
scaling of $\mu$ and $\sigma$ with $X_t$ (as shown in Fig.~\ref{fig.2}), 
we propose as an improvement of Eq.~(\ref{eq.sde1})~\cite{WeronBook}
\begin{equation}\label{eq.sde2}
  dX_t = \mu_t X_t dt + (a_t X_t+b_t) dL_t,
\end{equation}
where $L_t$ is an $\alpha$-stable L\'evy process, 
analogous to the Wiener process, except that the distribution of 
$dL_t$ follows a L\'evy-stable distribution with index $\alpha$, asymmetry $1$, 
location parameter $0$ and scale $1$ (using parametrization $1$ of 
Ref.~\cite{nolan2012stable}).
A cutoff in the noise term is added as above to ensure $dX_t\ge0$, so 
$\langle d L_t \rangle > 0$ and $\langle\Delta X_t \rangle$ is not given 
alone by the deterministic term $\mu_t X_t$ (even if $d L_t$ is understood in 
the Ito sense, as we do here \cite{oksendal2003stochastic}).
The parameters $\alpha,\mu,a,$ and $b$ depend on time~$t$ ($b_t$ is important 
only for small $X_t$ and $t$). Table~\ref{tab.SM1} summarizes all models.

\begin{table}
\begin{center}
\begin{tabular}{lllll}
\hline
\hline
Name & $\mathbb{P}(\Delta X_t|X_t)$ 
functional form & Parameters \\
\hline
LN & Lognormal & $\mu_t$, $\sigma_t$ \\
CEV & CEV & $\mu_t$, $\sigma_t$, $\beta_t$ \\
S & Lévy-stable & $\alpha_t$, $\mu_t$  $a_t$, $b_t$ \\
\hline
\hline
\end{tabular}
\end{center}
\caption{Sumsmary of models, see App.~\ref{app.1} for details.
\label{tab.SM1}}
\end{table}

\section{Improved data analysis}
We now discuss how to determine the parameters of the two models derived 
from Eq.~(\ref{eq.sde1}) and of the alternative model in Eq.~(\ref{eq.sde2}) 
and to test which model best describes the data.
The likelihood $\mathcal{L}_t$ of the models, for a fixed day $t$, 
is the product of the likelihoods of each distribution of 
$\Delta X_t$ conditioned on 
$X_t$ with respect to the parameters of the model $\theta$ as
\begin{equation}\label{eq.ml}
  \ln \mathcal{L}_t = \sum_{X_t} 
\sum_{\Delta X_t} 
N(\Delta X_t,X_t) \ln 
f(\Delta X_t | \theta, 
X_t),
\end{equation}
where $N(\Delta X_t,X_t)$ is the observed 
number of videos with a given 
$\Delta X_t,X_t$, 
and $f$ is the probability density function proposed by the models.
The best parameters $\theta$ are obtained maximizing $\ln \mathcal{L}_t$\footnote{This is performed numerically by minimizing 
$-\log\mathcal{L}$ through the Nelder-Mead algorithm, 
implemented in the Python package \texttt{scipy} as \texttt{optimize.fmin}.
We report the minimum value of $50$ repetitions performed with different (random) initial 
conditions (to avoid local minima).}
and the models are compared based on their (maximum) Likelihood, penalizing the 
addition of parameters (using the BIC\cite{noteBIC}).
The distributions $f$ we test are the same as above: Lognormal (LN) and 
Constant Elasticity of Variance (CEV), obtained from Eq.~(\ref{eq.sde1}), and 
L\'evy-stable (Stable), obtained from Eq.~(\ref{eq.sde2}).  
The latter is the $f$ resulting from Eq.~(\ref{eq.sde2}) because, 
since it seems to be a better fit to the data, we consider in this model that 
the time step $\Delta t$ is small such that $\Delta X_t \approx dX_t$, making 
the distribution to be fitted exactly the L\'evy.
Each of the distributions $f$ has different parameters that depend on $\theta$ 
and $X_t$, as summarized in Tab.~\ref{tab.SM1} and detailed in App.~\ref{app.1}.
Our approach based on Eq.~(\ref{eq.ml}) considers all conditional distributions $\mathbb{P}(\Delta 
X_t|X_t)$, avoiding the difficulties and arbitrary choices involved in the 
grouping of data in windows\footnote{There are two problems with the application of maximum-likelihood estimations to $\mathbb{P}(dX_t\mid X_t)$ for specific $X_t$:
(i) The estimation is quite sensitive to fluctuations, often failing to the determine
the tail exponent if no threshold is set~\cite{Newman2009} (many histograms with similar tails lead to very different values of $\alpha$).
(ii) Once the parameters of the fits for each 
$\mathbb{P}(dX_t\mid X_t)$ are extracted, it is not clear how to obtain the 
best parametrization of them, because of the presence of Lévy-distributed 
errors; the main limitations here are the truncation of fluctuations at 
$\Delta X_t=0$ 
(not to be confused with the lower positive limit on $X_t$ of Scheme 1 
described in App.~\ref{app.1}), and the discretization of the 
$\Delta X_t$ variable, 
which prevents a fit on the rescaled variable 
$z=(\Delta X_t-\mu(X_t))/\sigma(X_t)$. } as done in previous estimations and in 
Fig.~\ref{fig.2}.

\begin{figure}[htbp]
\begin{center}    
\includegraphics[width=0.98\columnwidth]{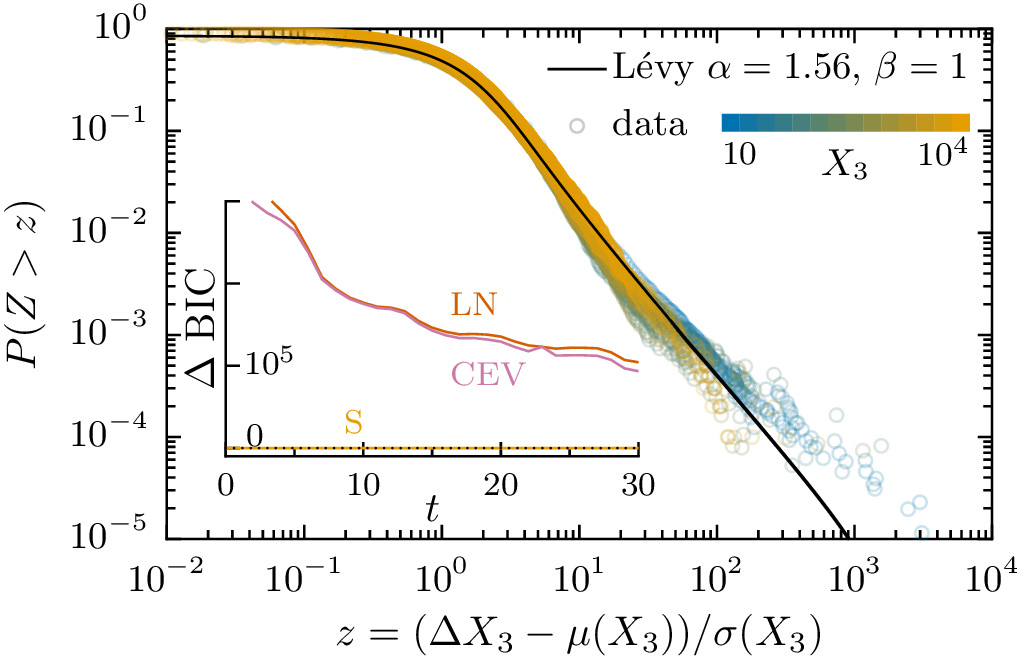}
\caption{
  Agreement of the model with respect to data.
  Main panel: complementary cumulative distribution of the views rescaled by 
  the fitted parameters for $t=3$.
  The rescaled histograms $\mathbb{P}(\Delta X_3|X_3)$ are plotted as 
  points, where each color corresponds to a different value of $X_3$; 
  the black line is the L\'evy-stable distribution 
  with location $0$ and scale $1$.
  Inset: BIC difference with respect to the S model\cite{noteBIC}.
  }
\label{fig.3}
\end{center}
\end{figure}

The application of the analysis described above to the YouTube data leads to significant evidence in favor of the L\'evy-stable model,
Eq.~(\ref{eq.sde2}). Figure~\ref{fig.3} shows how the this model allows for the collapse of the many $\mathbb{P}(\Delta X_t|X_t)$ in a single curve, well described by a L\'evy stable curve.
More formally, the BIC~\cite{noteBIC} difference of the Stable model with 
respect to the other models is above $10^5$ for all $0 \leq t \le 30$ 
(inset of Fig.~\ref{fig.3}), indicating very strong statistical support for our 
model.

\subsection{Dependence of parameters with respect to $t$}

\begin{figure}[htbp]
\begin{center}
\includegraphics[width=0.98\columnwidth]{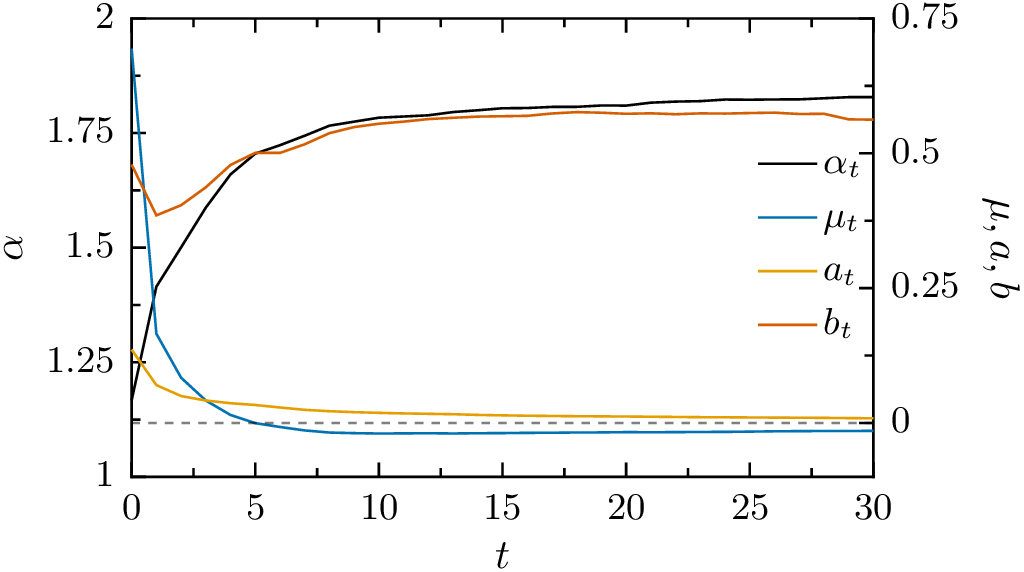}
\caption{
  Evolution of the parameters for the S model, Eq.~(\ref{eq.sde2}), in the 
first $30$ days after the release.
}
\label{fig.SM2}
\end{center}
\end{figure}

The parameters of the S model (Eq.~\ref{eq.sde2}) are explicitly time 
dependent, so we repeat the previous procedure for each of the days considered.
In Fig.~\ref{fig.SM2} the values of these are shown for the first 30 days 
after the publication of the videos.
The parameters show a strong dependence in $t$ in the the first week. 
In particular, $\mu_t$ decays in this period (reflecting a decay in 
the gain of views) and  $\alpha_t \approx 1.75$ for $t>5$.
It is worth to be noted is that the value of $\mu_t$ becomes negative; 
while apparently in contradiction with the positive slope of $\langle \Delta 
X_t| X_t\rangle$ (see Fig.~\ref{fig.2}(a)), it has to be recalled that the 
distribution is truncated at $\Delta X_t=0$.
The values of the averages from the data can be recovered through an exact, 
numerical computation.

If wanted, a model of the temporal dependence of  $\alpha_t,\mu_t,a_t,b_t$ can 
be introduced; in that case, it is possible to sum the likelihoods in 
Eq.~(\ref{eq.ml}) over $t$ and therefore to reduce the number of parameter of 
the models by avoiding independent fittings for each $t$.

Altogether, these analysis support our proposal of stochastic
differential equation with L\'evy noise, Eq.~(\ref{eq.sde2}), to 
describe the dynamics of popularity in YouTube.

\section{Prediction of big hits}
We now focus on the estimation of the probability of an item becoming 
a big hit after a given time. 
We define as a big hit at time $t$ the top $q\%$ videos with highest 
$X_t$ ($X_t > x_{t}^q$). 
We are particularly interested in estimating the probability 
$P(X_t>x_{t}^q| X_{t_0}=x_0)$ of videos that are not big hits at time 
$t_0<t$ (i.e., $x_0<x_{t_0}^{q})$ becoming big hits at time $t$. 
This probability quantifies how unpredictable the system is. 
For instance, in a deterministic (proportional growth) model, 
the rank of the videos does not change and therefore such 
probability is zero. A positive probability is thus a measure of the 
deviation of such perfect predictability.

\begin{figure}[htbp]
\begin{center}
\includegraphics[width=0.98\columnwidth]{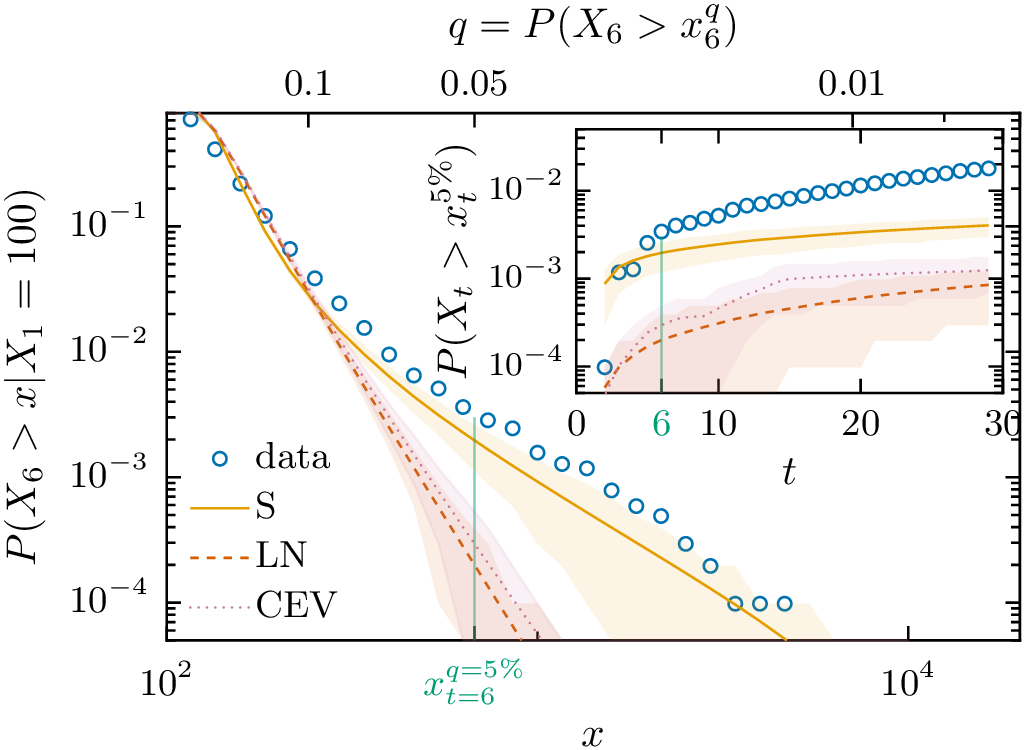}
\caption{
  Probability of videos becoming a big hit.
  Performance of the models evolved in time with respect to data;
  the selected videos had $100$ views $1$ day after their publication.
  Main panel: amount of videos that exceed a threshold $x$ at $t=6$.
  In the top axis, the quantiles $q$ are indicated.
  Inset: amount of videos that enter into the $5\%$ most viewed.
  Shaded areas: 95\% confidence intervals, by bootstrapping.
  }
\label{fig.4}
\end{center}
\end{figure}

As an example, we select the videos that had 100 views one day after 
publication, $X_1=100$, which belong to a rank of $q \approx 15\%$. 
We are interested in the probability of these videos having $X_t \gg 100$ at 
$t>1$.
To obtain the expectations of the models, we computed $P(X_t|X_1=100)$ 
iteratively from $P(dX_s|X_s)$ for $s=1, \ldots t$, using $X_1=100$ and the 
$t$-dependent parameters estimated in the previous section.
The results shown in Fig.~\ref{fig.4}(main panel) for $t=6$ confirm that the 
L\'evy-stable model predicts a substantially higher probability for large $X_t$ 
than alternative models.
In order to investigate the temporal dependence, we focus on the
probability of the videos improving their rank and being by day 
$t$ in the top $q=5\%$, using the previously computed probabilities
from the models and the thresholds $x_t^q$ estimated from data.
The results are summarized in Fig.~\ref{fig.4}(inset) and show that the 
L\'evy-stable model succeeds in estimating this probability in the short-term, 
while for the long-term the data shows an even higher probability (mixing of 
ranks).
The other models assign a video a substantially lower possibility of becoming a 
big hit, an effect of their highly predictable dynamics.
The fact that our model provides a good account for short-time intervals but 
not in the long run suggests the existence of correlations in the attribution 
of views that span multiple days and that are not accounted by our assumption 
of 
an independent noise. 
\section{Discussion and Conclusion}

Our finding that the growth of views in YouTube is governed by both linear
proportional growth and L\'evy fluctuations has important consequences for the
mathematical modeling of complex systems. 
First, it shows that, even if proportional growth is present, it cannot be 
attributed as the responsible for the origin of the heavy tails because
this is a feature already present in the fluctuations. 
Second, the use of Gaussian-based stochastic equations, such as 
Eq.~(\ref{eq.sde1}) or traditional Fokker-Planck equations, 
overestimate the predictability of videos, by neglecting the mobility of 
popularity.
We showed that better results are obtained in YouTube using a stochastic 
equation with L\'evy noise, Eq.~(\ref{eq.sde2}),  an approach that has been 
previously used in Physics~\cite{WeronBook}, climate 
research~\cite{ditlevsen1999observation}, and 
finance~\cite{mandelbrot1963prices}.
Our work indicates that this formalism, and possibly also kinetic
equations of the fractional type~\cite{metzler1999anomalous,brockmann2002levy}, 
should be considered in problems involving the dynamics of social-media items 
and, more generally, in models of the economy of attention.

Our results bring new insights on the attention economy of the Internet. 
The fact that the multiple factors affecting the popularity of videos can be 
effectively modeled by a L\'evy-stable distribution shows that the decision of 
different individuals are correlated to each other and lead itself to strong 
fluctuations. 
The L\'evy-stable distribution is invariant under convolution, i.e. if 
$X_1$,$X_2$ are stable, also $X_1+X_2$ is stable, and therefore it may naturally appears when multiple processes with diverging moments are combined (e.g., bursty activity patterns that characterize online social 
media).
One challenge for future work is to identify mechanistic models of the 
spreading of information on the Internet (e.g., models in which viral items 
spread through a social network) that are compatible with these 
fluctuations~\cite{watts2002simple,
maillart2008empirical,ratkiewicz2010characterizing}.
The presented analysis of fluctuations are enabled by the large availability of 
data in YouTube videos and we expect similar results to hold also in more 
general systems in which items compete for the attention of users.

\appendix

\section{Models}\label{app.1}

We compare the data collected with the distribution predicted by a series 
of simple models: from Eq.~(\ref{eq.sde1}), we derive the Lognormal (LN) and Constant Elasticity of Variance (CEV) models; from Eq.~(\ref{eq.sde2}) we derive the L\'evy-stable model (S). 
To compare with data, we compute the distributions of $X_{t+1}$ of the different
models. Note that for the LN and CEV models, these distributions are the 
result of integrating Eq.~(\ref{eq.sde1}) over a period of one day, while for the S models, 
this integration is not performed, i.e. we assume that in the period of one day 
the distribution of $X_{t+1}$ is essentially the one of the noise.

\subsection{Lognormal (LN)}

The LN model is defined by considering a linear scaling of the noise term in
Eq.~(\ref{eq.sde1})
\begin{equation} \label{eq.LN}
  dX_t = \mu_t X_t dt + \sigma_t X_t dW_t
\end{equation}
We integrate this equation for a time equal to 1 day (where we consider $\mu_t$ 
and $\sigma_t$ constant), such that $X_{t+1}$ is distributed lognormally, with 
a probability density function
\begin{widetext}
\begin{equation}
\mathbb{P}(X_{t+1}=x|X_{t}=x_0) = \frac{\exp\left(-(\log x - (\log x_0 + 
\mu_t-\sigma_t^2/2))^2/(2\sigma_t^2)\right)}{\sqrt{2\pi}\sigma_t x}
\end{equation}
\end{widetext}
$\Delta X_{t+1}=X_{t+1}-X_t$ is distributed 
also lognormally, since $X_t$ is fixed, 
but a truncation at 0 is necessary. Since the data is distributed on the 
natural numbers, we discretize as well the distribution, normalizing by the sum 
of the PDF over its new domain.

\subsection{Constant Elasticity of Variance (CEV)}
If instead of Eq.~(\ref{eq.LN}) the equation
\begin{equation} \label{eq.CEV}
  dX_t = \mu_t X_t dt + \sigma_t X_t^{\beta} dW_t
\end{equation}
is used, we have to use the distribution of the Constant Elasticity of 
Variance process (CEV), described in Ref.~\cite{chen2010cev}.
When $\beta<1$, it has the form
\begin{widetext}
\begin{equation}  
\mathbb{P}(X_{t+1}=x|X_{t}=x_0) = 
2(1-\beta)k^{\frac1{2(1-\beta)}}\left(xz^{1-4\beta}\right)^{\frac1{4(1-\beta)}}
e^{-x-z} I_{\left|\frac1{2(1-\beta)}\right|}\left(2\sqrt{xz}\right)
\end{equation}
\end{widetext}
with
\begin{align*}
  k &= \frac{\mu}{\sigma^2 (1-\beta)(e^{2\mu(1-\beta)}-1)} \\
  x &= k (x_0e^{\mu})^{2(1-\beta)} \\
  z &= k x^{2(1-\beta)}
\end{align*}
where $I$ is the modified Bessel function of the first kind.
The expression simplifies using the substitution $p=2(1-\beta)$:
\begin{equation}  
\mathbb{P}(X_{t+1}=x|X_{t}=x_0) = 
pk^{\frac1{p}}\left(xz^{2p-3}\right)^{\frac1{2p}}
e^{-x-z} I_{\left|\frac1{p}\right|}\left(2\sqrt{xz}\right)
\end{equation}
with
\begin{align*}
  k &= \frac{2\mu}{\sigma^2 p(e^{\mu p}-1)} \\
  x &= k x_0^p e^{\mu p} \\
  z &= k x^p
\end{align*}

When $\beta>1$, the distribution is the same as above 
but multiplied by $-1$. Note that the $\beta$ parameter is the 
exponent of the power-law tail that the distribution has asymptotically.
Here we also subtract $X_t$ to obtain a distribution of 
$\Delta X_t$,
which we also truncate, discretize, and normalize.

\subsection{Lévy-stable (S)}
In the S-model, defined in Eq.~(\ref{eq.sde2}),
$dX_t$ is Lévy-stable distributed with location parameter 
$m=\mu_t x_0$,
scale parameter $s = a_t x_0 + b_t$, asymmetry $\beta_L=1$ 
and its 
tail decays as an $\alpha$ power of $dX_t$\footnote{
++In our exploratory analysis we considered also two variants of the S-model in Eq.~(\ref{eq.sde2}): one where $b_t=0$ and one where the deterministic part is $\mu_t X_t + c_t$. A BIC analysis shows that the quality decreases significantly in the first case, but only marginally in the second case.
}.
These parameters correspond to the parametrization $1$ of 
Ref.~\cite{nolan2012stable}, where the characteristic function of $dX$ 
(there is no explicit form of the Lévy probability distribution function), 
$\phi_{dX}(k)$ is given by
\begin{widetext}
\begin{equation}
\log\phi_{dX}(k) = 
\begin{cases}
im k-s^{\alpha}|k|^{\alpha} 
\left[1+i\beta_L\,\tan\left(\frac{\pi\alpha}2\right)\text{sign}(k)\right] & 
\quad \alpha\neq1 \\
im k - s|k|\left[1+i\beta_L\frac2\pi \text{sign}(k) \log(|k|)\right] & 
\quad \alpha=1
\end{cases}
\end{equation}
\end{widetext}
We consider $\Delta X_t\approx dX_t$, and the parameters absorb the 
dependence on $\Delta t$.
In order to get the distribution $\mathbb{P}(X_{t+1}=x|X_{t}=x_0)$, the 
characteristic function has to be transformed to the real space, translated on 
the $X_{t+1}$ axis by an amount $x_0$, and then truncated at $X_{t+1}=x_0$, 
discretized, and normalized.

Numerically, the Lévy distribution is computed as:
\begin{itemize}
\item [(i)] the characteristic function (its Fourier transform) is inverted 
  numerically on a grid  in$\alpha\in(0.5,2)$ and $\beta\in(0,1)$ (with a 
  resolution of $0.05$, values of $\alpha$  below $0.5$ are very unlikely and 
  for $\beta<0$ the distribution can be computed from the one of $-\beta$ using 
  symmetry);
\item [(ii)] for general values $\alpha, \beta$, we compute the distribution 
  as an interpolation of the values on the grid (using the Catmull-Rom cubic 
  splines).
\item[(iii)] the numerical integration often becomes unstable in the tails 
  of the distribution (large $x$). In order to avoid this problem, we use the 
  power-law approximation described in Ref.~\cite{newman2001clustering} to describe 
  the distribution beyond a threshold.  
\end{itemize}
We provide the code of this procedure in the package 
PyLevy~\cite{miotto2016pylevy}. 
It contains routines to compute the PDF of the Lévy distribution and to fit it.

\section{Fluctuations expected from existing models}\label{app.2}

Here we discuss the form that 
$\mathbb{P}(\Delta X_t|X_t)$ has for the 
classical
linear proportional growth models. There are basically two schemes of
implementing linear proportional growth in order to get heavy-tailed distributions:
\begin{itemize}
\item Scheme 1: Champernowne \cite{champernowne1953model} introduces a lower 
  positive limit to $X$. A master equation is defined to regulate 
  the transitions to different states (amount of views), which eventually leads 
  to a stationary distribution with a power-law tail.
  This argument was formalized and popularized in 
  Refs.~\cite{kesten1973random,gabaix1999zipf}, using a linear stochastic differential equation 
  (as Geometric Brownian Motion, GBM) which, in the limit of long time, converges to 
  a heavy tailed distribution; note that in this scheme, all items start with 
  the same initial condition.
\item Scheme 2: Yule and Simon \cite{yule1925mathematical,simon1955class} 
  design a scheme where views are added to different items while at the same
  time new items are introduced, resulting in a power law distribution.
  This is basically the model known as 
  \emph{preferential attachment}~\cite{barabasi1999emergence} in the context of 
  network growth.
  Here the items start in different conditions, since the system is growing in 
  the number of items, hence the first ones are privileged.
\end{itemize}

We focus on the transition probability 
$\mathbb{P}(\Delta X_t|X_t)$.
Scheme 1 (GBM) is a mechanism that leads to a heavy-tailed distribution 
asymptotically, but also relies on the possibility of negative growth rates 
$dX_t$, which is not realistic in the context of videos' views, and more 
generally in the context of cumulative allocation of attention.
For an infinitesimal increase of time ($\Delta t\approx dt$), 
the distribution is Normal, but
if a finite time interval is considered ($\Delta t\gg dt$), the 
integration of the model results
in a Lognormal distribution.

The Scheme 2, instead, is fundamentally different and can be thought of as a
Polya Urn process where at a given time $t_0$, a number of views $N$ is
assigned to a set of videos $M$ that has exactly $x_0$ views already.
The probability of assigning a view to a particular video is, of course,
proportional to the amount of views each video has.
The distribution 
$\mathbb{P}(\Delta X_t|X_t)$ of views among 
the videos is of the 
Beta Binomial type, with means
\begin{equation}
  \mathbb{E}_{BB}(\Delta X_t|X_t=x_0) = 
\frac{N}{M}
\end{equation}
and variance
\begin{equation}
  \mathbb{V}_{BB}(\Delta X_t|X_t=x_0) = 
\frac{N(M-1)(Mx_0+N)}{M^2(Mx_0+1)}
\end{equation}
and can be roughly approximated by a Normal distribution with same mean and 
variance.
The variance scales in two regimes because of the $M x_0+N$ term: when 
$x_0\lesssim N/M$, $\sigma\propto x_0^{-1/2}N/M$, and when $x_0\gtrsim N/M$, 
$\sigma\propto\sqrt{N/M}$.
Notice though, that the amount of views allocated, $N$, is not independent.
In fact, since growth is linear, we expect $N\propto x_0M$, so we have 
in this particular case $\sigma\propto x_0^{1/2}$.

In conclusion, for Scheme 1, we expect normal or Lognormal distributions of 
$\Delta X_t$, depending on the choice of 
whether integrating over a finite time or 
not, while for Scheme 2 we expect approximately normal distributions, 
with variance scaling as $X_t$.

\begin{acknowledgments}
We thank I. Sokolov for helpful discussions.
\end{acknowledgments}


\end{document}